\documentclass[unrestricted]{stagstatus}
\usepackage{hyperref}
\usepackage{blindtext}
\usepackage[most]{tcolorbox}
\usepackage{graphicx}
\usepackage{listings}

\usepackage{amsmath}
\usepackage{amssymb}
\usepackage{amsthm}

\usepackage[simplified]{pgf-umlcd}
\usepackage{tikzit}

\tikzstyle{basic}=[fill=white, draw=black, shape=circle]
\tikzstyle{square}=[fill=white, draw=black, shape=rectangle]
\tikzstyle{big dashed}=[fill=white, draw=black, shape=circle, minimum width=1cm, dashed]
\tikzstyle{vertical ellipse dashed}=[fill=none, draw=blue, minimum width=0.75cm, minimum height=3cm, ellipse, dashed, tikzit shape=rectangle, tikzit draw=blue, tikzit fill=white]
\tikzstyle{small vertical ellipse dashed}=[fill=none, draw=blue, shape=circle, tikzit fill=white, tikzit draw=blue, dashed, minimum width=0.75cm, minimum height=1.5cm, tikzit shape=rectangle, ellipse]
\tikzstyle{tiny vertical ellipse dashed}=[fill=none, draw=blue, shape=circle, tikzit fill=white, ellipse, dashed, minimum width=0.75cm, minimum height=1cm, tikzit shape=rectangle]
\tikzstyle{red}=[fill={rgb,255: red,191; green,0; blue,64}, draw=black, shape=circle]
\tikzstyle{green}=[fill={rgb,255: red,0; green,128; blue,128}, draw=black, shape=circle]
\tikzstyle{blue}=[fill=blue, draw=black, shape=circle]
\tikzstyle{huge dashed}=[fill=white, draw=black, shape=circle, dashed, minimum width=2cm]
\tikzstyle{medium}=[fill=white, draw=black, shape=circle, minimum width=1cm]
\tikzstyle{pale green}=[fill={rgb,255: red,173; green,231; blue,0}, draw=black, shape=circle, minimum width=1cm]
\tikzstyle{horizontal ellipse dashed}=[fill=white, draw=black, tikzit draw=magenta, tikzit shape=rectangle, minimum width=3cm, minimum height=0.75cm, ellipse, dashed]
\tikzstyle{minsize}=[fill=white, draw=black, shape=circle, minimum width=0.75cm]
\tikzstyle{horizontal ellipse green}=[fill={rgb,255: red,191; green,255; blue,0}, draw=black, tikzit draw={rgb,255: red,191; green,255; blue,0}, tikzit shape=rectangle, minimum width=3cm, minimum height=0.75cm, ellipse, dashed]
\tikzstyle{horizontal ellipse blue}=[fill={rgb,255: red,107; green,203; blue,255}, draw=black, tikzit draw=blue, tikzit shape=rectangle, minimum width=3cm, minimum height=0.75cm, ellipse, dashed]
\tikzstyle{smallblack}=[fill=black, draw=black, shape=circle, inner sep=0 pt, minimum size=3 pt]
\tikzstyle{smallSquare}=[fill=white, draw=black, shape=rectangle, inner sep=0 pt, minimum size=6 pt]
\tikzstyle{smallCircle}=[fill=white, draw=black, shape=circle, inner sep=0 pt, minimum size=6 pt]
\tikzstyle{big vertical ellipse dashed}=[fill=none, draw=blue, shape=circle, tikzit shape=rectangle, ellipse, dashed, minimum width=0.95cm, minimum height=3.7cm]
\tikzstyle{smallred}=[fill={rgb,255: red,191; green,0; blue,64}, draw=black, shape=circle, inner sep=0 pt, minimum size=6 pt]
\tikzstyle{smallblue}=[fill=blue, draw=blue, shape=circle, inner sep=0pt, minimum size=3pt]

\tikzstyle{directed}=[->]
\tikzstyle{undirected}=[-, line width=1pt]
\tikzstyle{directed red}=[draw=red, ->, line width=1pt]
\tikzstyle{directed green}=[draw={rgb,255: red,0; green,128; blue,128}, ->, line width=1pt]
\tikzstyle{directed blue}=[draw=blue, ->, line width=1pt]
\tikzstyle{directed purple}=[draw={rgb,255: red,128; green,0; blue,128}, ->, line width=1pt]
\tikzstyle{undirected red}=[-, draw=red, line width=1pt]
\tikzstyle{undirected green}=[-, draw={rgb,255: red,0; green,107; blue,61}, line width=1pt]
\tikzstyle{undirected blue}=[-, draw=blue, line width=1pt]
\tikzstyle{undirected purple}=[-, draw={rgb,255: red,128; green,0; blue,128}, line width=1pt]
\tikzstyle{undirected dashed}=[-, line width=1pt, dashed]
\tikzstyle{orange dashed}=[-, draw={rgb,255: red,255; green,128; blue,0}, dashed, line width=1.5pt]
\tikzstyle{directed dash}=[->, dashed]
\tikzstyle{blue dashed}=[-, draw=blue, dashed, line width=1pt]
\tikzstyle{green dashed}=[-, draw={rgb,255: red,0; green,162; blue,0}, dashed, line width=1pt]
\tikzstyle{blue filled}=[-, fill={blue!20}, draw=blue, line width=1pt, opacity=0.5, tikzit fill=white]
\tikzstyle{red filled}=[-, fill={red!20}, line width=1pt, draw=red, opacity=0.5, tikzit fill=white]
\tikzstyle{green filled}=[-, line width=1pt, draw={rgb,255: red,0; green,107; blue,61}, opacity=0.5, tikzit fill=white, fill={rgb,255: red,149; green,255; blue,179}]
\tikzstyle{orange filled}=[-, fill={orange!20}, draw=orange, line width=1pt, opacity=0.5, tikzit fill=white]
\tikzstyle{undirected dashed}=[-, draw=black, dashed, line width=1pt]

\usepackage{todonotes}
\setuptodonotes{color=blue!30, size=\tiny}
\definecolor{eggshell}{rgb}{0.94, 0.92, 0.84}

\definecolor{codegreen}{rgb}{0,0.6,0}
\definecolor{codegray}{rgb}{0.5,0.5,0.5}
\definecolor{codepurple}{rgb}{0.58,0,0.82}
\definecolor{backcolour}{rgb}{0.95,0.95,0.92}

\lstdefinestyle{mystyle}{
    backgroundcolor=\color{backcolour},   
    commentstyle=\color{codegreen},
    keywordstyle=\color{magenta},
    numberstyle=\tiny\color{codegray},
    stringstyle=\color{codepurple},
    basicstyle=\ttfamily\footnotesize,
    breakatwhitespace=false,         
    breaklines=true,                 
    captionpos=b,                    
    keepspaces=true,                 
    numbers=left,                    
    numbersep=5pt,                  
    showspaces=false,                
    showstringspaces=false,
    showtabs=false,                  
    tabsize=2
}

\title{Technical Report~(1)} 
\statusdate{\today}
\project{EPSRC Early Career Fellowship,  EP/T00729X/1.}

\begin{document}

\frontmatter

\setcounter{page}{1}

\section{Summary}
Spectral Toolkit of Algorithms for Graphs~(STAG) is an open-source C++ and Python library of efficient spectral algorithms for graphs.
Our objective is to implement advanced
graph algorithms developed through algorithmic spectral graph theory, while making it practical to end users.
This series of technical reports is to document our progress on STAG, including
implementation details, engineering considerations, 
and the data sets against which our implementation is tested.
The report is structured as follows:
\begin{itemize}
    \item Section~\ref{sec:local_clustering} describes the local clustering algorithm, which is the main update in this STAG release. The discussion is at a high level such that domain knowledge beyond basic algorithms is not needed.
    \item Section~\ref{sec:user_guide} provides a user guide to the essential features of STAG which allow a user to apply local clustering.
    \item Section~\ref{sec:examples} includes experiments and demonstrations of the functionality of STAG.
    \item Finally, Section~\ref{sec:technical} discusses several technical details; these include our choice of implemented algorithms, the default setup of parameters, and other technical choices. 
    We leave these details to the final section, as it's not necessary for the reader to understand this when using STAG.
\end{itemize}

\subsection{Implemented Algorithms}
STAG 1.2 provides an implementation of the following key algorithms.

\paragraph{Local Graph Clustering.}
Given a large graph and some starting vertex $v$ in the graph, the goal of local graph clustering is to find some cluster containing $v$. Moreover, the running time of the algorithm should depend only on the size of the returned cluster and should be independent of the total size of the graph~\cite{ST13}.

 STAG  provides the first open-source local clustering algorithm which does not require the entire graph to be 
loaded into memory.
This allows users to apply local clustering on massive graphs stored on disk
or even in a cloud database, such as Neo4j\footnote{\url{https://neo4j.com/}}. 
Section~\ref{sec:examples} demonstrates these applications.

\paragraph{Spectral Clustering.}
One of the most fundamental algorithms from Spectral Graph Theory is the spectral clustering algorithm~\cite{macgregorTighterAnalysisSpectral2022, ngSpectralClusteringAnalysis2001, vonluxburgTutorialSpectralClustering2007}.
Spectral clustering is ``global'' in the sense that it returns a partition of the entire vertex set of the graph.

\paragraph{Generating Graphs from Random Models.}
The Stochastic Block Model (SBM) and Erd\H{o}s-R\'{e}nyi model are popular random graph models which are frequently used to evaluate and analyse graph algorithms.
STAG provides several convenient methods to generate graphs from these models.

\section{Local Graph Clustering} \label{sec:local_clustering}
Graph clustering algorithms are designed to partition an input graph into two or more clusters.
As a basic technique in data science and machine learning, graph clustering has many applications in numerous areas of computer science and beyond.
Most graph clustering algorithms need to read an entire input graph for the clustering task, which is computationally expensive if the graph is massive.
If one is interested only in some ``local'' cluster information, then local graph clustering provides a more efficient method.

\begin{figure}[th]
    \centering
    \begin{tikzpicture}
	\begin{pgfonlayer}{nodelayer}
		\node [style=basic] (0) at (-3, 2.25) {};
		\node [style=basic] (1) at (-1, 2.25) {};
		\node [style=basic] (2) at (0, 0.5) {};
		\node [style=basic] (3) at (-1, -1.25) {};
		\node [style=basic] (4) at (-3, -1.25) {};
		\node [style=red] (5) at (-3.75, 0.5) {};
		\node [style=basic] (6) at (1.75, 2) {};
		\node [style=basic] (7) at (2.75, 3) {};
		\node [style=basic] (8) at (3, 0.5) {};
		\node [style=basic] (9) at (2.75, -2) {};
		\node [style=basic] (10) at (1.75, -1) {};
		\node [style=basic] (11) at (1.25, 0.5) {};
		\node [style=none] (14) at (4.25, 2.75) {};
		\node [style=none] (15) at (4.25, -1.75) {};
		\node [style=none] (16) at (4.5, 0.5) {};
		\node [style=none] (17) at (-4.8, 2.5) {};
		\node [style=none] (18) at (-4.3, 3) {};
		\node [style=none] (19) at (0.2, 3) {};
		\node [style=none] (20) at (0.7, 2.5) {};
		\node [style=none] (21) at (0.7, -1.5) {};
		\node [style=none] (22) at (0.2, -2) {};
		\node [style=none] (23) at (-4.3, -2) {};
		\node [style=none] (24) at (-4.8, -1.5) {};
		\node [style=none] (25) at (-4.2, 2.3) {$S$};
		\node [style=none] (26) at (-4.2, -0.1) {$v$};
	\end{pgfonlayer}
	\begin{pgfonlayer}{edgelayer}
		\draw [style=undirected] (0) to (1);
		\draw [style=undirected] (0) to (2);
		\draw [style=undirected] (0) to (4);
		\draw [style=undirected] (0) to (5);
		\draw [style=undirected] (1) to (2);
		\draw [style=undirected] (1) to (3);
		\draw [style=undirected] (1) to (4);
		\draw [style=undirected] (1) to (5);
		\draw [style=undirected] (2) to (3);
		\draw [style=undirected] (2) to (4);
		\draw [style=undirected] (3) to (4);
		\draw [style=undirected] (3) to (5);
		\draw [style=undirected] (4) to (5);
		\draw [style=undirected] (6) to (7);
		\draw [style=undirected] (6) to (8);
		\draw [style=undirected] (6) to (9);
		\draw [style=undirected] (6) to (10);
		\draw [style=undirected] (6) to (11);
		\draw [style=undirected] (7) to (8);
		\draw [style=undirected] (7) to (9);
		\draw [style=undirected] (7) to (10);
		\draw [style=undirected] (7) to (11);
		\draw [style=undirected] (8) to (9);
		\draw [style=undirected] (8) to (10);
		\draw [style=undirected] (8) to (11);
		\draw [style=undirected] (9) to (10);
		\draw [style=undirected] (9) to (11);
		\draw [style=undirected] (10) to (11);
		\draw [style=undirected dashed] (7) to (14.center);
		\draw [style=undirected dashed] (8) to (16.center);
		\draw [style=undirected dashed] (9) to (15.center);
		\draw [style=undirected] (1) to (6);
		\draw [style=undirected] (3) to (10);
		\draw [style=blue dashed] (24.center)
			 to (17.center)
			 to [bend left=45, looseness=1.50] (18.center)
			 to (19.center)
			 to [bend left=45, looseness=1.75] (20.center)
			 to (21.center)
			 to [bend left=45, looseness=1.50] (22.center)
			 to (23.center)
			 to [bend right=315, looseness=1.50] cycle;
	\end{pgfonlayer}
\end{tikzpicture}
    \caption{Given a massive graph containing some small cluster $S$, a local clustering algorithm takes as input a vertex $v \in S$ and returns an approximation of $S$ without exploring the whole graph.
    \label{fig:local_clustering}}
\end{figure}
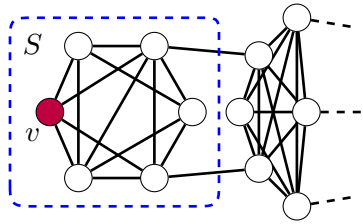

Typically, the objective of local graph clustering is to find some highly-connected vertex set~(cluster) in an input graph.
Let's assume that $S$ is a highly-connected vertex set  of an  underlying undirected graph $G$, i.e., $S$ forms a cluster. Then, a local clustering algorithm is given some vertex $v\in S$ as input, and returns some set $S'$ such that $S'$ is a reasonable approximation of the target set $S$.
Moreover, the running time of the algorithm is proportional to the size of $S$ and independent of the size of $G$. 
In applications, local clustering can be viewed as a search for related objects: given a query vertex, a local clustering algorithm returns a set of closely related vertices.
Figure~\ref{fig:local_clustering} illustrates local graph clustering.

Andersen, Chung, and Lang~\cite{ACL} introduced a key local clustering algorithm which we refer to as the ACL algorithm.
At a high level, the algorithm finds a local cluster by analysing the behaviour of random walks on the graph, beginning at the starting vertex.
The ACL algorithm has proved extremely useful and has inspired extensive further research and applications~\cite{andersenLocalPartitioningDirected2007, MS21, takaiHypergraphClusteringBased2020, yinLocalHigherorderGraph2017}.
The  \lstinline{local_cluster} method of STAG provides an implementation of the ACL local clustering algorithm.

Existing open-source local clustering methods require that the entire graph is loaded into RAM in order to apply the local algorithm.
In this sense, they are not truly ``local'' since they cannot be applied to graphs larger than the available memory and the total running time depends on the size of the graph.
STAG provides the first open-source local clustering algorithm which can be applied to massive graphs without loading them into RAM.
Moreover, the provided interface is simple and the algorithm can be
applied to graphs stored in memory, on disk, or in a Neo4j database.

\section{User's Guide to Local Graph Clustering with STAG} \label{sec:user_guide}
This section provides a guide to the essential features of  STAG  which allow a user to apply local clustering.
Section~\ref{subsec:installation} describes how to install the STAG C++ and STAG Python libraries.
Then, Section~\ref{subsec:file_formats} introduces the graph file formats supported by STAG and demonstrates the methods for reading and writing graphs to disk.
Finally, Section~\ref{subsec:graph_classes} explains the graph classes provided by  STAG  and Section~\ref{subsec:local_clustering} documents the \lstinline{local_cluster} method for local clustering.

Although most of the examples in this section use C++, the functionality of STAG C++ is also available in STAG Python.
Appendix~\ref{app:ug_python} includes example code demonstrating how to perform local clustering with STAG Python.
The full documentation of  STAG C++ and STAG Python  is available on the STAG library website.

\subsection{Installation of STAG} \label{subsec:installation}
This section describes how to install STAG for use with C++ and Python.

\paragraph{Installing STAG for C++.}
STAG is built on the Eigen and Spectra C++ libraries, and these must be installed before STAG.
For information on installing Eigen and Spectra, please refer to their documentation.
For convenience, Appendix~\ref{app:dependencies} provides a bash script which, at the time of writing, will install Eigen and  Spectra on a standard Linux system.
Then, the latest version of  STAG  should be downloaded from
\begin{equation*}
    \mbox{\url{https://github.com/staglibrary/stag/releases}.}
\end{equation*}
After downloading and extracting the source code,  STAG   can be compiled and installed with \lstinline{cmake}.
\begin{lstlisting}[language=bash]
    mkdir build_dir
    cd build_dir
    cmake ..
    sudo make install
\end{lstlisting}
Once STAG has been installed, it is available for use in C++ projects built with the \lstinline{cmake} build tools.
The following \lstinline{cmake} code will link a C++ project with  STAG.
\begin{lstlisting}
    find_package(stag REQUIRED)
    include_directories(${STAG_INCLUDE_DIRS})
    target_link_libraries(YOUR_PROJECT stag)
\end{lstlisting}
An example STAG project demonstrating the full \lstinline{cmake} configuration is available at 
\begin{equation*}
\mbox{\url{https://github.com/staglibrary/example-stag-project}.}
\end{equation*}

 \paragraph{Installing STAG for Python.}
  STAG Python  can be installed from the Python Package Index with the pip tool.
\begin{lstlisting}[language=bash]
    python -m pip install stag
\end{lstlisting}
 Then, the modules of  STAG   can be directly imported into any Python script.

\subsection{File Formats} \label{subsec:file_formats}
 STAG  supports two simple file formats for storing graphs on disk: EdgeList and AdjacencyList.
Many graph datasets are provided in EdgeList format~\cite{snapnets}, and will work directly with STAG.

\paragraph{EdgeList File Format.}
In an EdgeList file, each line corresponds to one edge in the graph.
A line consists of two integer node IDs and an optional edge weight, all separated with spaces.
Here is an example of a simple EdgeList file.
\begin{lstlisting}
  # This is a comment
  0 1 0.5
  1 2 1
  2 0 3
\end{lstlisting}
In this example, line 2 defines an edge between nodes $0$ and $1$ with weight $0.5$.

\paragraph{AdjacencyList File Format.}
In an AdjacencyList file, each line corresponds to one node in the graph.
A line consists of the node ID, followed by a list of adjacent nodes.
The node IDs at the beginning of each line must be sorted in increasing order.
Here is an example of  a simple AdjacencyList file.
\begin{lstlisting}
  # This is a comment
  0: 1 2
  1: 0 3 2
  2: 0 1
  3: 1
\end{lstlisting}
In this example, node $1$ has edges to nodes $0$, $2$, and $3$.

\paragraph{Working with Files.}
STAG provides several methods for reading, writing, and converting between EdgeList and AdjacencyList files, as demonstrated in the following example.
\begin{lstlisting}[language=C++]
  #include <stag/graphio.h>
  ...
    // Read an AdjacencyList graph
    std::string filename = "mygraph.adjacencylist";
    stag::Graph myGraph = stag::load_adjacencylist(filename);

    // Save an EdgeList graph
    std::string new_filename = "mygraph.edgelist";
    stag::save_edgelist(myGraph, new_filename);

    // Convert an AdjacencyList to EdgeList directly
    stag::adjacencylist_to_edgelist(filename, new_filename);
  ...
\end{lstlisting}
\subsection{Graph Classes} \label{subsec:graph_classes}
\begin{figure}[ht]
    \centering
    \input{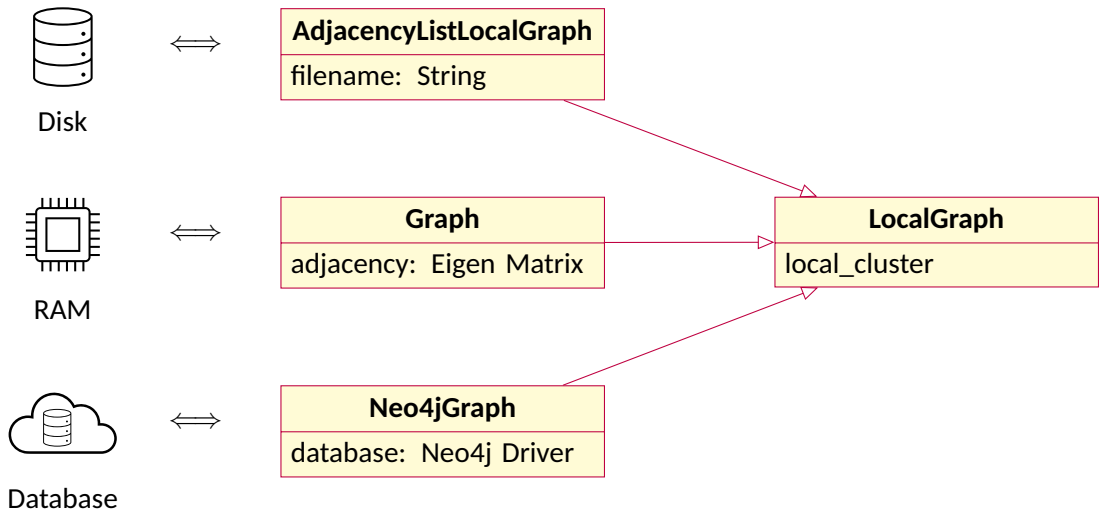}
    \caption{The graph classes provided by STAG. Every class inherits from the abstract \lstinline{LocalGraph} class, which provides the local clustering method.
    The key difference between the different classes is the location of the graph data. The \lstinline{AdjacencyListLocalGraph} class reads node adjacency data from disk,
    the \lstinline{Graph} class stores the entire graph in RAM, and the \lstinline{Neo4jGraph} class queries node adjacency data from a Neo4j database.
    }
    \label{fig:classes}
\end{figure}

STAG provides several graph classes which can be applied for a wide variety of applications.
Figure~\ref{fig:classes} summarises the available graph classes.

\paragraph{The LocalGraph Class.}
STAG provides an abstract \lstinline{stag::LocalGraph} class which defines the data structure necessary to apply local clustering.
The only required method on the data structure is \lstinline{neighbors(v)} which returns a list of the neighbors of node $v$.
Every graph class provided by STAG inherits from LocalGraph.

\paragraph{The Graph Class.}
The \lstinline{stag::Graph} class is the basic graph object within the STAG library.
The class stores the adjacency matrix of the graph in memory as a sparse matrix.

\paragraph{The AdjacencyListLocalGraph Class.}
The \lstinline{stag::AdjacencyListLocalGraph} class provides an implementation of the \lstinline{stag::LocalGraph} interface for a graph stored on disk as an AdjacencyList.
The graph is loaded into memory in a local way only.
This allows for local algorithms to be executed on very large graphs stored on disk without loading the whole graph into memory.
The following example demonstrates how to create an \lstinline{AdjacencyListLocalGraph} with STAG C++.
\begin{lstlisting}[language=C++]
  #include <stag/graph.h>
  ...
    // Create an AdjacencyListLocalGraph
    std::string filename = "mygraph.adjacencylist";
    stag::AdjacencyListLocalGraph myGraph(filename);

    // Get the neighbours of node 0
    std::vector<long long> neighbors = myGraph.neighbors_unweighted(0);
  ...
\end{lstlisting}

\paragraph{The Neo4jGraph Class.}
 STAG Python  additionally provides the \lstinline{Neo4jGraph} class, which provides an implementation of the \lstinline{LocalGraph} interface for a graph stored in a Neo4j database.
The following example shows how to create the \lstinline{Neo4jGraph} using the database connection information.
\begin{lstlisting}[language=Python]
  import stag.neo4j

  # Connect to the database
  uri = "<Database URI>"
  username = "neo4j"
  password = "<password>"
  my_graph = stag.neo4j.Neo4jGraph(uri, username, password)

  # Print the neighbors of node 0
  print(my_graph.neighbors_unweighted(0))
\end{lstlisting}
The \lstinline{Neo4jGraph} class provides additional methods for querying the properties of the nodes in the database.
The details are available in the full STAG documentation.

\subsection{Local Clustering} \label{subsec:local_clustering}
STAG provides the following \lstinline{local_cluster} method.
\begin{lstlisting}[language=C++]
  std::vector<long long> stag::local_cluster(stag::LocalGraph* graph,
                                             long long         seed_vertex,
                                             double            target_volume)
\end{lstlisting}
Given a graph and a starting vertex, the \lstinline{local_cluster} method finds a cluster close to the starting vertex.
The running time of the algorithm is proportional to the size of the returned cluster and independent of the size of the entire graph.
The parameters of the method are described as follows: 
\begin{itemize}
    \item \textbf{graph} - a \lstinline{LocalGraph} object. 
    This could be a \lstinline{Graph}, an \lstinline{AdjacencyListLocalGraph}, or a \lstinline{Neo4jGraph}.
    \item \textbf{seed\_vertex} - the starting vertex in the graph.
    \item \textbf{target\_volume} - an estimate of the volume of the target cluster.
    This parameter does \emph{not} impose a hard constraint on the algorithm and so an approximate volume is sufficient.
\end{itemize}
When working with very large graphs, it is recommended to use the \lstinline{AdjacencyListLocalGraph} object for local clustering in order to avoid the overhead of reading the entire graph into memory.
Section~\ref{sec:allg_example} demonstrates the advantage of using the \lstinline{AdjacencyListLocalGraph} for local clustering.
The following code demonstrates a complete program which uses STAG C++ to find a local cluster in a graph stored in an AdjacencyList file on disk.
\begin{lstlisting}[language=C++]
  #include <iostream>
  #include <stag/graph.h>
  #include <stag/cluster.h>

  int main() {
    // Create the graph backed by a large file on disk
    std::string filename = "mygraph.adjacencylist";
    stag::AdjacencyListLocalGraph mygraph(filename);

    // Perform local clustering
    int start_vertex = 1;
    double target_volume = 100;
    auto cluster = stag::local_cluster(&mygraph, start_vertex, target_volume);

    // Print the returned cluster
    for (auto v : cluster) std::cout << v << ", ";
    std::cout << std::endl;
  }
\end{lstlisting}

\section{Showcase studies} \label{sec:examples}
 STAG  makes local clustering  straightforward for a variety of applications, and this section presents some examples of  local clustering with STAG.
The code used to produce all experimental results is available at
\begin{equation*}
\mbox{\url{https://github.com/staglibrary/local-clustering-case-study}.}
\end{equation*}
All experiments are performed on an HP ZBook laptop with an 11th Gen Intel(R) Core(TM) i7-11800H @ 2.30GHz processor and 32 GB RAM.

\subsection{Example 1} \label{sec:allg_example}
The advantage of local clustering over other clustering algorithms is that the running time of local clustering is proportional to the size of the returned cluster and independent of the total size of the graph.
If we first load the entire graph into memory before applying local clustering, then we lose the advantage of the sub-linear running time.
For this reason, STAG provides the \lstinline{AdjacencyListLocalGraph} class which provides local access to a graph stored on disk without reading the entire graph.
In this example, we compare the running time of local clustering on an \lstinline{AdjacencyListLocalGraph} object, which accesses the graph locally on disk, and a \lstinline{Graph} object, which loads the entire graph into memory.

We generate graphs of various sizes from the stochastic block model as follows.
Given parameters $k$, $p$, and $q$, we create a graph with $k$ clusters $C_1, \ldots C_k$, each containing 1,000 vertices.
For every pair of vertices $(u, v) \in V \times V$, we add the edge $(u, v)$ with probability $p$ if $u$ and $v$ are in the same cluster and with probability $q$ otherwise. 
We always set $p = 0.01$ and $q = 0.001 / k$.
This ensures that the conductance of the constructed clusters is always close to $0.1$.

\begin{figure}[ht]
\centering
\includegraphics[width=0.45\columnwidth]{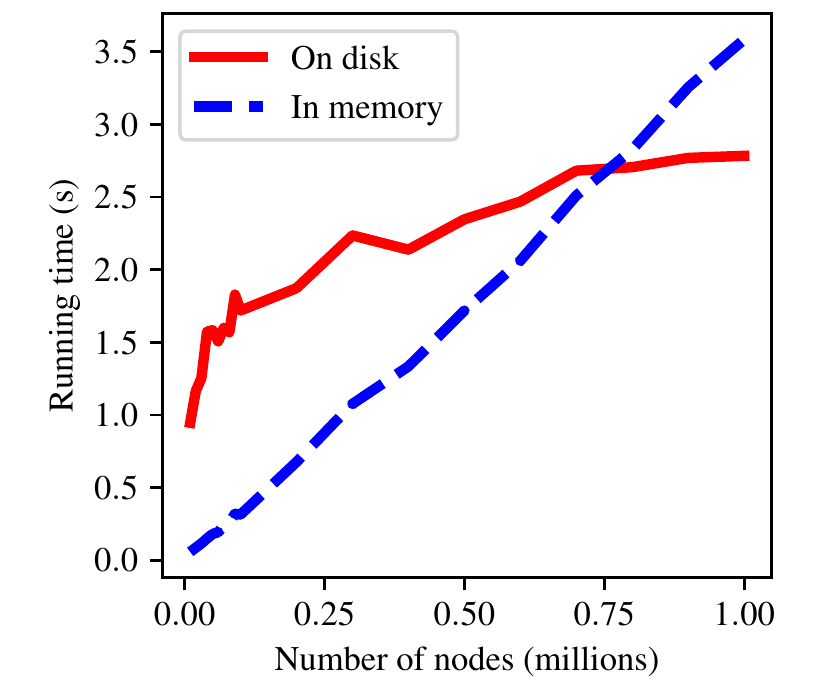}
\hspace{0.05\columnwidth}
\includegraphics[width=0.45\columnwidth]{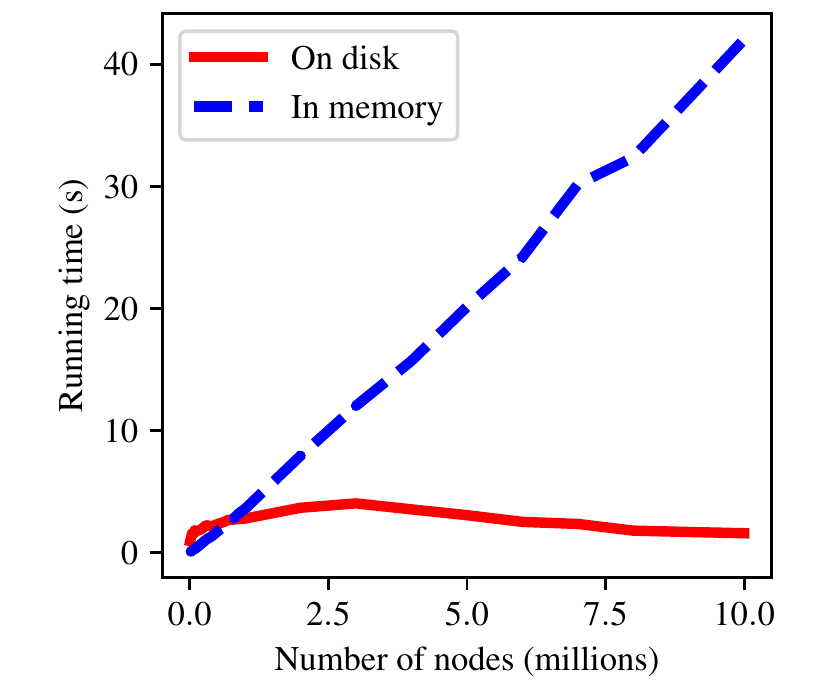}
\caption{The comparison of the running time of local clustering on graphs in memory and on disk. For massive graphs, reading the graph locally from disk is significantly faster.
\label{fig:sbm}}
\end{figure}

We perform local clustering on the constructed graphs for a random starting node and target volume 20,000, and compare the following two methods:
\begin{itemize}
    \item \textbf{In memory}: the entire graph is loaded into memory as a \lstinline{Graph} object before applying local clustering. 
    \item \textbf{On disk}: the graph is read locally from a file on disk with an \lstinline{AdjacencyListLocalGraph} object.
\end{itemize}
Figure~\ref{fig:sbm} shows the running time of the local clustering algorithm for each method across a range of graph sizes.
These results demonstrate that for large graphs, the overhead of reading the entire graph into memory dominates the running time of the algorithm and reading the graph directly from disk is significantly more efficient.

\subsection{Example 2}
In the second example, we demonstrate the applicability of local clustering for finding sets of related nodes in a real-world graph.
We use the \lstinline{wiki-topcats} dataset~\cite{yinLocalHigherorderGraph2017} which is a graph of Wikipedia hyperlinks constructed in 2011.
The graph includes the pages in the top 100 Wikipedia categories and includes 1,791,489 vertices and 28,511,807 edges.
The dataset is available on the SNAP datasets page~\cite{snapnets} as an EdgeList file.
We convert the EdgeList to an AdjacencyList with the \lstinline{edgelist_to_adjacencylist} method, and  use the \lstinline{AdjacencyListLocalGraph} object
for local clustering.

With a few lines of code,  STAG 
allows us to create a ``related pages'' search using local clustering.
By providing a search page and setting the target volume to be 100, the local clustering  returns a set of pages which are closely connected
to the search page.
Figure~\ref{fig:wiki_examples} shows some example of local clustering results from the Wikipedia graph.

\begin{figure} [ht]
\begin{tcbraster}[raster columns=3,raster rows=1, raster equal height]
\begin{tcolorbox}[colback={bgcolor},colupper=black,title=Search page: Emacs]
Emacs \\
Robert J. Chassell \\
Zmacs \\
Climacs \\
Aquamacs \\
Dunnet (game) \\
Bernard Greenberg \\
Macro recorder \\
GNU Manifesto \\
TNT (instant messenger) \\
Agda (theorem prover)
\end{tcolorbox}    
\begin{tcolorbox}[colback={bgcolor},colupper=black,title=Search page: Stag Hound]
Stag Hound \\
Over-canvassed sailing \\
Sea Serpent (clipper) \\
Medium Clipper \\
Donald McKay \\
Extreme clipper \\
Memnon (clipper) \\
Flying Cloud (clipper) \\
Eleanor Creesy \\
Weigh anchor \\
Stowage
\end{tcolorbox}    
\begin{tcolorbox}[colback={bgcolor},colupper=black,title=Search page: Stagflation]
Stagflation \\
Agflation \\
Biflation \\
Differential accumulation \\
Embedded liberalism \\
Supply shock \\
Shimshon Bichler \\
Douglas Harper \\
Online savings account \\
Jonathan Nitzan
\end{tcolorbox}    
\end{tcbraster}
    \caption{Example of local clusters found in the Wikipedia dataset.}
    \label{fig:wiki_examples}
\end{figure}

\subsection{Example 3}
In this example, we demonstrate local clustering on a Neo4j database in the cloud.
We first follow the Neo4j documentation to create a cloud database using the AuraDB service~\cite{neo4j}.
We use the ``Movies'' example dataset provided by Neo4j.
Then, by creating a \lstinline{Neo4jGraph} object with STAG Python, we are able to search for related movies using
local clustering.
Listing~\ref{lst:movie_code} shows the complete Python script used to perform this search, and
Figure~\ref{fig:movie_examples} shows some of the search results.
\begin{lstlisting}[language=Python, label={lst:movie_code}, caption=Python code for local clustering with a Neo4j database.]
import stag.neo4j
import stag.cluster

# Initialise the graph object with the Neo4j database credentials
uri = "<Database URI>"
username = "neo4j"
password = "<password>"
g = stag.neo4j.Neo4jGraph(uri, user, password)

# Perform local clustering
seed_id = g.query_id("title", "The Matrix")
cluster = stag.cluster.local_cluster(g, seed_id, 5)

# Print the names of the returned movies.
for node_id in cluster:
  title = g.query_property(node_id, 'title')
  print(title)
\end{lstlisting}

\begin{figure}[ht]
\begin{tcbraster}[raster columns=3,raster rows=1, raster equal height]
\begin{tcolorbox}[colback={bgcolor},colupper=black,title=Search: The Matrix]
The Matrix \\
The Matrix Reloaded \\
The Matrix Revolutions \\
V for Vendetta \\
Ninja Assassin
\end{tcolorbox}    
\begin{tcolorbox}[colback={bgcolor},colupper=black,title=Search: You've Got Mail]
You've Got Mail \\
RescueDawn \\
When Harry Met Sally \\ 
As Good as It Gets \\
Sleepless in Seattle \\
Joe Versus the Volcano
\end{tcolorbox}    
\begin{tcolorbox}[colback={bgcolor},colupper=black,title=Search: A Few Good Men]
A Few Good Men \\
Hoffa \\
As Good as It Gets \\
Stand By Me \\
\end{tcolorbox}    
\end{tcbraster}
    \caption{Example of local clusters found in the Neo4j movies dataset.}
    \label{fig:movie_examples}
\end{figure}
\section{Technical Considerations} \label{sec:technical}
In this section we discuss the technical choices made in the design and implementation of STAG.
We discuss our choice of implemented algorithm, the setting of the default parameters,
and the implementation of the \lstinline{AdjacencyListLocalGraph} class for reading a graph locally from a file on disk.

\subsection{Implemented Algorithm}
Many algorithms have been proposed for local clustering, including those based on PageRank~\cite{ACL, andersenLocalPartitioningDirected2007}, the evolving set process~\cite{andersenAlmostOptimalLocal2016, andersenFindingSparseCuts2009}, and network flows~\cite{fountoulakisPNormFlowDiffusion2020}.
We chose to implement the algorithm based on PageRank presented by Andersen, Chung, and Lang~\cite{ACL}, and we refer to this as the ACL algorithm.
We chose this algorithm because it is relatively simple, easy to understand, and effective in practice.
Furthermore, the theoretical guarantees for the ACL algorithm are optimal up to constant factors.\footnote{The original analysis by Andersen et al.~\cite{ACL} has an extra factor of $\log(n)$ in the approximation guarantee. This factor is not necessary and has been removed in later analysis using the same technique~\cite{MS21, takaiHypergraphClusteringBased2020}.}
The ACL algorithm requires two parameters:
\begin{itemize}
    \item the $\alpha$ parameter controls the ``teleport probability'' of the personalised PageRank; and
    \item the $\epsilon$ parameter controls the approximation error of the approximate PageRank calculation.
\end{itemize}
STAG provides the \lstinline{local_cluster_acl} method which allows the user to specify the parameters $\alpha$ and $\epsilon$ directly.
\begin{lstlisting}[language=C++]
  std::vector<long long> stag::local_cluster_acl(stag::LocalGraph* graph,
                                                 long long         seed_vertex,
                                                 double            alpha,
                                                 double            epsilon)
\end{lstlisting}
For convenience, STAG also provides the \lstinline{local_cluster} method which requires only an estimate of the volume of the target cluster.
Given a volume $\gamma$, the \lstinline{local_cluster} method uses the parameters $\alpha = 1/2000$ and $\epsilon = 1 / (20 \gamma)$ for the ACL algorithm.

\subsection{Reading Graphs Locally From Disk}
A key feature of  STAG  is the \lstinline{AdjacencyListLocalGraph} class which reads the neighbourhood information of a graph in a local way from an AdjacencyList file on disk.
Since the data in an AdjacencyList file is sorted according to the node ID, we can query the neighbors of any node in $O(\log(n))$ time by binary search of the AdjacencyList file.
As demonstrated in Section~\ref{sec:allg_example}, this additional logarithmic factor in the running time is much preferable to the cost of reading the entire graph into memory when applying local algorithms to massive graphs.

\bibliographystyle{plain}
\bibliography{reference}

\appendix

\section{Installing STAG Dependencies} \label{app:dependencies}
For convenience, we provide the following bash script for installing the STAG C++ dependencies.
At the time of writing, this will download and install the Eigen and Spectra libraries.
\begin{lstlisting}[language=bash]
  # Create a directory to work in
  mkdir libraries
  cd libraries
 
  # Install Eigen
  wget https://gitlab.com/libeigen/eigen/-/archive/3.4.0/eigen-3.4.0.tar.gz
  tar xzvf eigen-3.4.0.tar.gz
  cd eigen-3.4.0
  mkdir build_dir
  cd build_dir
  cmake ..
  sudo make install
  cd ../..
 
  # Install Spectra
  wget https://github.com/yixuan/spectra/archive/v1.0.1.tar.gz
  tar xzvf v1.0.1.tar.gz
  cd spectra-1.0.1
  mkdir build_dir
  cd build_dir
  cmake ..
  sudo make install
  cd ../.. 
\end{lstlisting}

\section{User Guide Examples using STAG Python} \label{app:ug_python}
This section includes example code omitted from Section~\ref{sec:user_guide}
demonstrating how to use STAG Python for local clustering.

\subsection{Working with Files}
The following example demonstrates how to read and write AdjacencyList and EdgeList files with STAG Python.
\begin{lstlisting}[language=Python]
  import stag.graphio

  # Read an AdjacencyList graph
  filename = "mygraph.adjacencylist"
  myGraph = stag.graphio.load_adjacencylist(filename)

  # Save an EdgeList graph
  new_filename = "mygraph.edgelist"
  stag.graphio.save_edgelist(myGraph, new_filename)

  # Convert an AdjacencyList to EdgeList directly
  stag.graphio.adjacencylist_to_edgelist(filename, new_filename)
\end{lstlisting}

\subsection{Graph Classes}
The following example shows how to create an \lstinline{AdjacencyListLocalGraph} object with STAG Python.
\begin{lstlisting}[language=Python]
  import stag.graph

  # Create the graph object
  filename = "mygraph.adjacencylist"
  my_graph = stag.graph.AdjacencyListLocalGraph(filename)

  # Show the neighbors of node 0
  print(my_graph.neighbors_unweighted(0))
\end{lstlisting}

\subsection{Local Clustering}
The following example gives a complete program for finding a local cluster in a graph stored in an AdjacencyList file with STAG Python.
\begin{lstlisting}[language=Python]
  import stag.graph
  import stag.cluster

  # Create the graph object
  filename = "mygraph.adjacencylist"
  mygraph = stag.graph.AdjacencyListLocalGraph(filename)

  # Find a local cluster
  start_vertex = 1
  target_volume = 100
  cluster = stag.cluster.local_cluster(mygraph, start_vertex, target_volume)

  # Display the result
  print(cluster)
\end{lstlisting}

\end{document}